\documentclass[preprints,review,accept,moreauthors,pdftex]{Definitions/mdpi} 

\firstpage{1} 
\pubvolume{xx}
\issuenum{1}
\articlenumber{5}
\pubyear{2019}
\copyrightyear{2019}
\history{Received: date; Accepted: date; Published: date}

\usepackage[utf8]{inputenc}
\usepackage{amsmath}
\usepackage{amsfonts}
\usepackage{amssymb}
\usepackage{graphicx}

\newcommand{\dd}{\mathrm{d}}
\newcommand{\cobj}[1]{\texttt{\textcolor{blue}{#1}}}

\Title{Selected Topics in Numerical Methods for Cosmology}

\Author{Sandro Dias Pinto Vitenti $^{1,\dagger}$\orcidA{} and Mariana Penna-Lima $^{1,\dagger}$\orcidB{}}

\AuthorNames{Sandro Dias Pinto Vitenti and Mariana Penna-Lima}

\address[1]{%
$^{1}$ \quad Universidade de Brasília, Instituto de Física, Caixa Postal 04455, Brasília, DF, 70919-970, Brazil}

\corres{Correspondence: pennalima@unb.br}

\firstnote{These authors contributed equally to this work.}

\abstract{The large amount of cosmological data already available (and in the near future) makes necessary the development of efficient numerical codes. Many software products have been implemented to perform cosmological analyses considering one or few probes. The need of multi-task software is rapidly increasing, in order to combine numerous cosmological probes along with their specificity (e.g., astrophysical descriptions and systematic errors).  In this work we mention some of these libraries, bringing out some challenges they will face in the few-percent error era (on the cosmological parameters).  We review some concepts of the standard cosmological model, and examine some specific topics on their implementation,  bringing, for example,  the discussion on how some quantities are numerically defined in different codes.  We also consider implementation differences between public codes, mentioning their advantages/disadvantages. }


\begin{document}

\section{Introduction}
Cosmology is dedicated to study the evolution of the Universe since its primordial
epoch until today. The description and modeling of its different stages and
observables are, in general, divided into two classes: deterministic and
stochastic predictions. The standard model of cosmology is described by the
General Relativity (GR) theory of gravitation with a homogeneous and isotropic
background metric, which provides deterministic predictions for some
cosmological observables. On top of this background an inhomogeneous description
is laid, usually through a perturbation theory, for which the predictions are of
stochastic nature. For instance, the distances are among the
cosmological/astrophysical observables related just with the background. In
contrast, the perturbations describe all the deviations from a homogeneous and
isotropic universe, for this reason, were it deterministic, it would have to
describe all astrophysical object positions and properties in a given scale.
Such deterministic description for perturbations is unfeasible and consequently
a probabilistic approach is chosen. This means that instead of describing the
exact positions and properties of each astrophysical object in a given scale, we
model the probability distribution of these quantities. For example, when
describing the mass distribution in the universe, we use the matter power
spectrum $P_m$, which is a simple way to express the (Gaussian) probability
distribution of finding a given matter contrast function $\delta_m(x) \equiv
\delta\rho_m(x) / \rho_m$ ($\delta\rho_m$ denotes the perturbation of the
matter energy density around its mean value $\rho_m$).

Furthermore, the initial conditions for the very early universe are given in
terms of the quantization of its initial perturbations (e.g. in an inflationary
scenario, for a review see~\cite{Linde2017}). So, apart from the problem of
decoherence (the evolution from quantum to classical probability distributions,
for example see~\cite{Martin2018}), the initial conditions are already given in
terms of stochastic distributions. Nevertheless,  observations provide
positions and characteristics of actual astrophysical objects as well as the
errors in these determinations. Consequently, we have two sources of
indeterminacies, the probabilistic aspect of the observation errors and the
stochastic nature of our modeling.

In practice we have two classes of problems in numerical cosmology. The first being related to background observables, such as the distance modulus of type Ia supernovae (SNeIa), cosmic
chronometers and $H_0$ determination. These depend on quantities determined solely
by the background model. For this reason, the numerical methods involved are
simple and usually require solutions of a small set of Ordinary Differential
Equations (ODEs). Once the background is computed, the comparison with data is
done by taking into consideration the statistical errors in the observables'
determination. Any observable in this class is extremely useful since they
depend only on background modeling and, consequently, do not rely on the more
sophisticated (but more complex) description of the inhomogeneous universe. The
second class is composed by the observables related to the inhomogeneities. For
example, Cosmic Microwave Background (CMB) radiation, Large Scale Structure
(LSS) observables, such as galaxy spatial correlation, galaxy cluster counts and
correlation, and gravitational lensing. For some of these observables, e.g. CMB, the
linear perturbation theory is enough (up to a given scale) to compute the
observable statistical distribution, whereas for others, e.g. galaxy spatial
correlation, the perturbation theory must be corrected for scales beyond its
validity.

In the last two decades Cosmology has became a data-driven science thanks to the
large amount of high-quality observational data that has been released
\cite{Hinshaw2013, SDSS-IIICollaboration, PlanckCollaboration2015, DES2018}. The
complexity of the theoretical model of the observables along with their
astrophysical features requires sophisticated numerical and statistical tools.
In this chapter, we will discuss some of the numerical and statistical methods
that have been largely used in Cosmology. Most of these methods can be found in
the Numerical Cosmology (\texttt{NumCosmo}) library \cite{Vitenti2012c}, which
is a free software written in C on top of the
\texttt{GObject}\footnote{\url{https://developer.gnome.org/gobject/stable/}} and
\texttt{GObject Introspection}\footnote{\url{https://developer.gnome.org/gi/stable/}}
tools.\footnote{Briefly speaking, \texttt{GObject} provides an object-oriented
	framework for C programs, while \texttt{GObject Introspection} generates
	bindings for other languages, such as \texttt{Python}, \texttt{Perl} and
	\texttt{JavaScript}.} 
Popular codes such as
CAMB (\url{https://camb.info/})~\cite{Lewis2000, Howlett2012a, Lewis} and CLASS
(\url{http://class-code.net/})~\cite{Lesgourgues2011, Blas2011,
	Lesgourgues2011a, Lesgourgues2011b} are used to compute both background and
perturbation (linear and non-linear) quantities. Their statistical counterparts,
respectively, CosmoMC~\cite{Lewis2002, Lewis2013} and
MontePython~\cite{Audren2013}, implement not only the observable likelihoods but
the tools necessary to explore the parameter space of the underlying models
mainly through Markov Chain Monte Carlo (MCMC) method. In
Sec~\ref{sec:background} we describe the numerical approaches to solve the
background model and the observables related to it. In Sec~\ref{sec:lin_pert} we introduce the
tools to solve the linear perturbation problem along with the related observable quantities. At last, in Sec~\ref{sec:can} we discuss the most common cancellation errors found in numerical cosmology.
 
\section{Numerical Methods}	

In the next sections we will briefly outline the theoretical aspects of the background model and then move to the discussion on how to solve them numerically.

\subsection{Background}
\label{sec:background}

In this section we will describe the different ways to compute observables
related to the background cosmology, for this reason we introduce the necessary
concepts as we go, for a detailed discussion on these see for
example~\cite{Weinberg2008, Peter2009}. The majority of the cosmological models
relies on a metric theory. In particular, we consider the homogeneous and
isotropic metric, also known as the Friedmann-Lema\^{\i}tre-Robertson-Walker
metric (FLRW),
\begin{equation}
\label{eq:rw_metric} 
ds^2 = -c^2\,dt^2 + a^2 (t) \left [ dr^2 + S_K^2(r) (d\theta^2 + \sin^2 \theta  d\phi^2) \right ]\;,
\end{equation} 
where $c$ is the speed of light and 
\begin{equation}
S_K(r) = 
\begin{cases}
r & K = 0, \\
{\sin\left(r\sqrt{K}\right)}/{\sqrt{K}}, & K > 0,\\
{\sinh\left(r\sqrt{\vert K \vert}\right)}/{\sqrt{\vert K \vert}}, & K < 0,
\end{cases}
\end{equation}
for respectively flat, spherical and hyperbolic spatial hyper-surfaces. This
metric has just one function to determine, the scale factor $a(t)$, and the
Gaussian curvature $K$. In practice (which is the case for all cited softwares)
we use instead of $K$, the curvature density parameter at the present day
$\Omega_K^0 \equiv -Kc^2/(a_0H_0)^2$, where the super and subscript  $0$ indicate the quantity today, the Hubble parameter
$H_0 = \dot{a}/a \vert_0$ and $\dot{}$ denotes the derivative with respect to the
cosmic time $t$. Note that theoretical work usually define $K \in \{-1,\; 0,\;
1\}$ (using a given unit, such as $\mathrm{Mpc}^{-2}$) generating an one-to-one
relation between $a_0$ and $\Omega_K^0$, while in numerical codes, such as CAMB,
CLASS and NumCosmo, it is habitual to let $a_0$ as an arbitrary value defined by
the user and use the definition above to determine $K$ through $\Omega_K^0$.

In practice, we do not measure $a(t)$ directly, but related quantities such as
the distances to astronomical objects. Considering a null trajectory of photons
emitted by a galaxy traveling along the radial direction to us, we have that
\begin{equation}
r = c \int_{t_e}^{t_0} \frac{\dd t}{a(t)},
\end{equation}
where $t_e$ and $t_0$ are the emitted and observed times (here $t_0$ represent
the present time). Note that $r$ refers to the distance at the comoving
hyper-surface, i.e., the hyper-surface where $a(t_c) = 1$ [see
Eq.~\eqref{eq:rw_metric}]. This means that $r$ is the distance between the point
of departure and arrival both projected (through the Hubble flow) back to the
comoving hyper-surface. In other words, using the coordinates in
Eq.~\eqref{eq:rw_metric}, we define a foliation of the space-time where each
sheet is defined by $t=\mathrm{constant}$, then the photon leaves the source at $t_e$
at the point $p_e$ within the $t_e$ hyper-surface and arrives at point $p_0$ at
the hyper-surface $t_0$. Since $p_e$ and $p_0$ are defined at different
hyper-surfaces we need to transport them to the same hyper-surface in order to
have a meaningful definition of spatial distance between them. Then we project both
points $p_e$ and $p_0$ at the comoving hyper-surface by following the Hubble
flow that passes through these points back to the hyper-surface $t_c$
[$a(t_c)=1$] where $r$ gives the spatial distance between them.

In the literature it is often defined $a_0 = 1$ (for the flat case), that is, the comoving hyper-surface coincides with today's hyper-surface and $r$ would provide the distance between points today. However, as discussed above in the numerical codes one needs a general definition for the whole parameter space (in this case for any value of $\Omega_K$). For this reason, we keep $a_0$ arbitrary and define the comoving distance projected back to today's hyper-surface as 
\begin{equation}
\label{eq:dist_com}
d_c(z) \equiv a_0 r = d_H \int_0^z \frac{\dd \bar{z}}{E(\bar{z})},
\end{equation}
where the Hubble radius is $d_H = c / H_0$, $E(z) = H(z) / H_0$ is the normalized Hubble function, and the distance is now written in terms of the observable $z$ that is the redshift, 
\begin{equation}
1 + z \equiv \frac{a_0}{a(t)}.
\end{equation}
Finally, similar to the comoving distance we introduce the comoving time
\begin{equation}\label{ct}
\eta = c\int_0^t\dd \bar{t} \frac{1}{a(\bar{t})},
\end{equation}
where we chose the conformal time such that $\eta(t=0) = 0$. The only differences between the comoving and the conformal distances are a factor of $a_0$ and the integration limits.

\subsection{Time variables}

When solving the background it is usually useful to choose which variable
to use as time. This is particularly important if the evolution of the
perturbations will be also computed, since it requires the evaluation of
several background observables for each time step in the perturbations'
integration.

We first note that there is an one-to-one relation between $z$, $a$ and
$d_c$  (both $a_0$ and $t_0$ are fixed quantities).\footnote{The relation
	between $z$ and $d_c$ is monotonic since $E>0$.} The relation between
these variables and the cosmic time requires the choice of a reference
hypersurface to anchor the time definition. For instance it is common to
choose $t = 0$ to mark the singularity of the classical FLRW metric,
i.e., $a(0) = 0$. This choice has some drawbacks. In order to compute
$t(z)$ or $\eta(z)$, for example, we need to evaluate
\begin{equation}\label{eq:t:eta}
t = \int_z^\infty\dd \bar{z}\frac{1}{(1+\bar{z})H(\bar{z})} \quad \text{and} \quad \eta = \frac{1}{a_0}\int_z^\infty\dd \bar{z}\frac{1}{	H(\bar{z})}.
\end{equation}
However, the above integrals include the whole history of the background
metric, from $t = 0$ ($z=\infty$) to $z$. This means that, in practice, we
need to know the values of $H(z)$ for this whole range, including the matter-dominated phase, radiation phase, inflationary or bounce phase, etc. The available
numerical codes include the computation of $t(z)$ and/or $\eta(z)$. But one
should remark that these quantities are obtained from integrals as above and extrapolating the radiation phase till the singularity, i.e., they ignore any inflationary or bounce phases and assume a radiation dominated singularity in the past.

Both NumCosmo and CAMB compute the times $t$ and $\eta$ integrating the Hubble function $H(z)$ and extrapolating the radiation era till the singularity. CLASS uses the approximation of $\eta(z) \approx
(1+z)/(a_0H) = 1/(aH)$ which result from assuming $H(z) \propto (1+z)^2$
for $z$ in the radiation era. This means that there will probably be small shifts
between the $\eta$ variable when comparing different codes, furthermore, the choice of precision (and technique) used in the integration may also result in small differences. 

Another option for time variable, used by different objects in NumCosmo, is the scale factor $a$ or its logarithm $\alpha \equiv -\ln(a)$.\footnote{The minus sign in the definition of $\alpha$ is included in order to $\alpha$ monotonically increases with $t$ or $\eta$.} Of course, such variable cannot be used when the Hubble function $H$ change signs (for example in a bounce). But this variable is useful when describing the expansion phase since it is possible to obtain an analytic expression for $H(a)$ in different scenarios. In these cases we do not need to solve the integrals~\eqref{eq:t:eta} to relate $H$ and the time variable.

\subsection{Distances}
\label{sec:dist}

The base of several cosmological observers are the cosmological distances. Besides the comoving distance [Eq.~\eqref{eq:dist_com}] other useful distances are:
the transverse comoving $d_M(z)$, the angular diameter $d_A(z)$ and luminosity distances $d_L(z)$, respectively defined as~\cite{Hogg2007}:
\begin{eqnarray}
d_M = \left\{
\begin{array}{c l}	
d_H \frac{1}{\sqrt{\Omega_K}} \sinh \left( \sqrt{\Omega_K} d_c/d_H  \right) & \text{for} \quad \Omega_K > 0 \\
d_c & \text{for} \quad \Omega_K = 0 \\
d_H \frac{1}{\sqrt{\vert \Omega_K \vert }} \sin \left( \sqrt{\vert \Omega_K \vert} d_c/d_H  \right) & \text{for} \quad \Omega_K < 0,
\end{array}\right.
\end{eqnarray}  
\begin{equation}
d_A(z) = \frac{d_M(z)}{(1 + z)}  \quad \text{and} \quad d_L(z) = (1 + z) d_M(z). 
\end{equation}

In its turn, the Hubble function $H(z)$ can be recovered using observational
data, which allows the model-independent approaches (see \cite{Vitenti2015} and references therein) besides the usual modeling  assuming a  theory of Gravitation,
i.e., independent of the equations of motion discussed in the last section.

\begin{figure}
\begin{center}
	\includegraphics[scale=0.5]{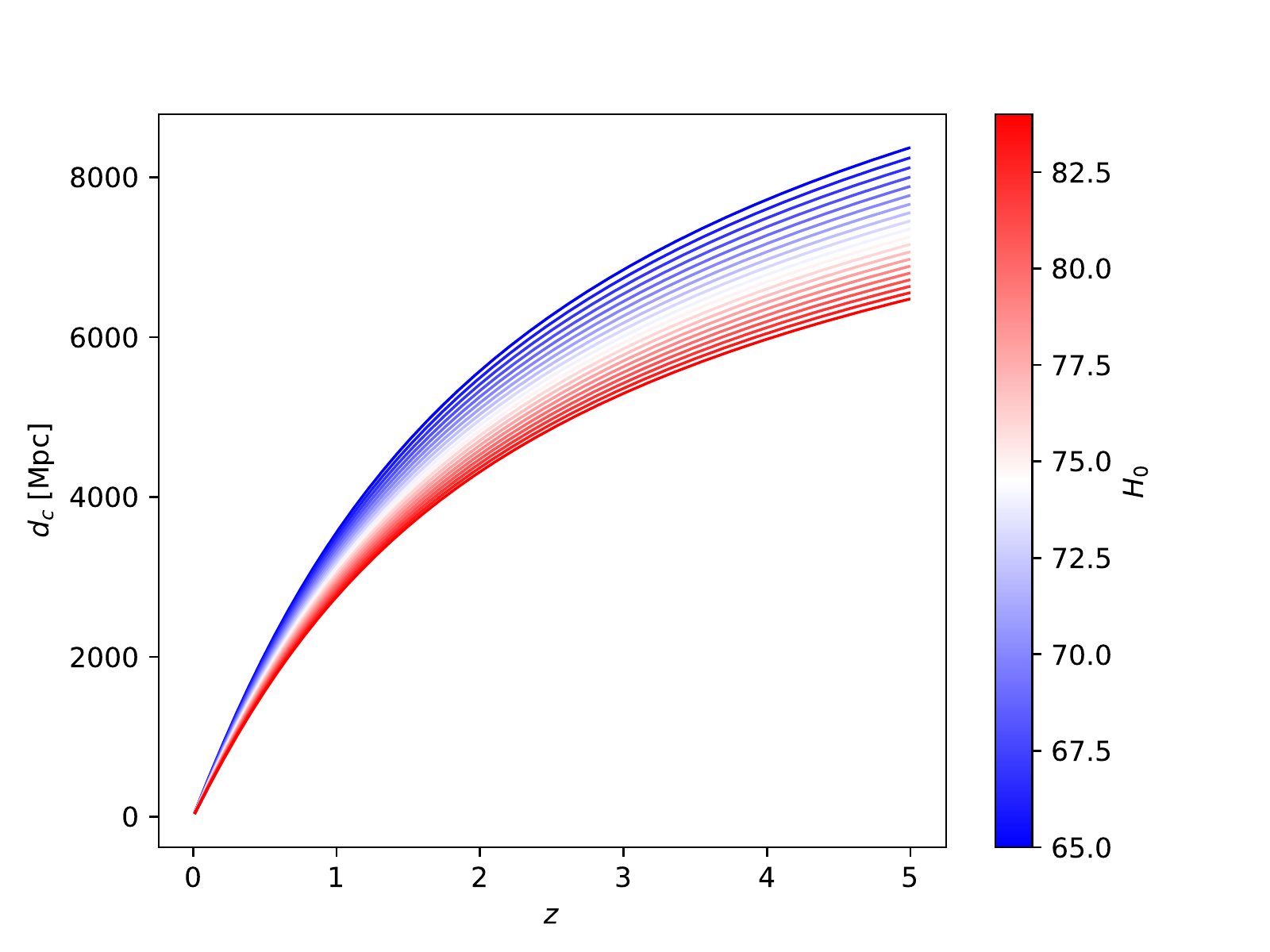}
	\includegraphics[scale=0.5]{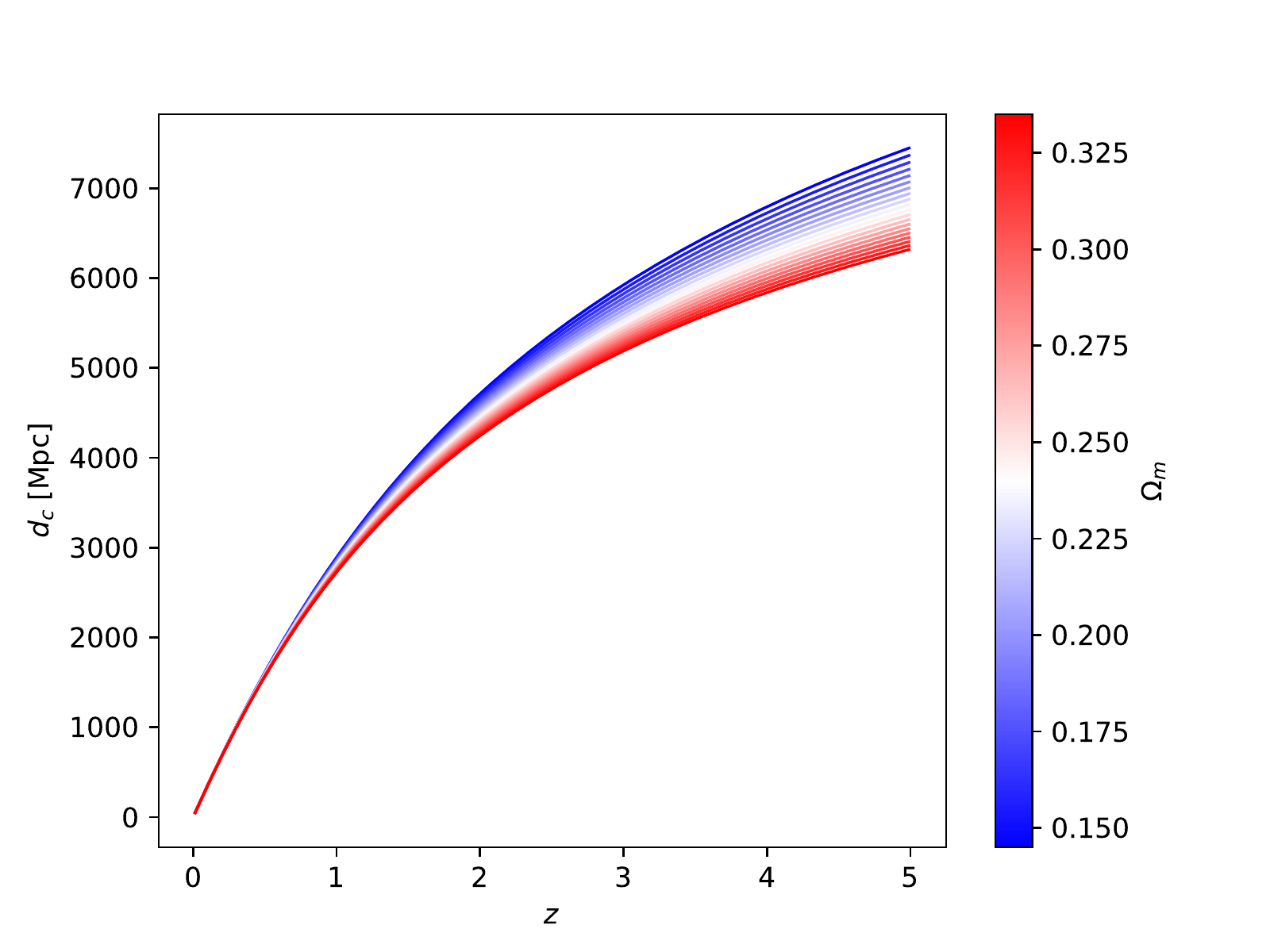}
	\includegraphics[scale=0.5]{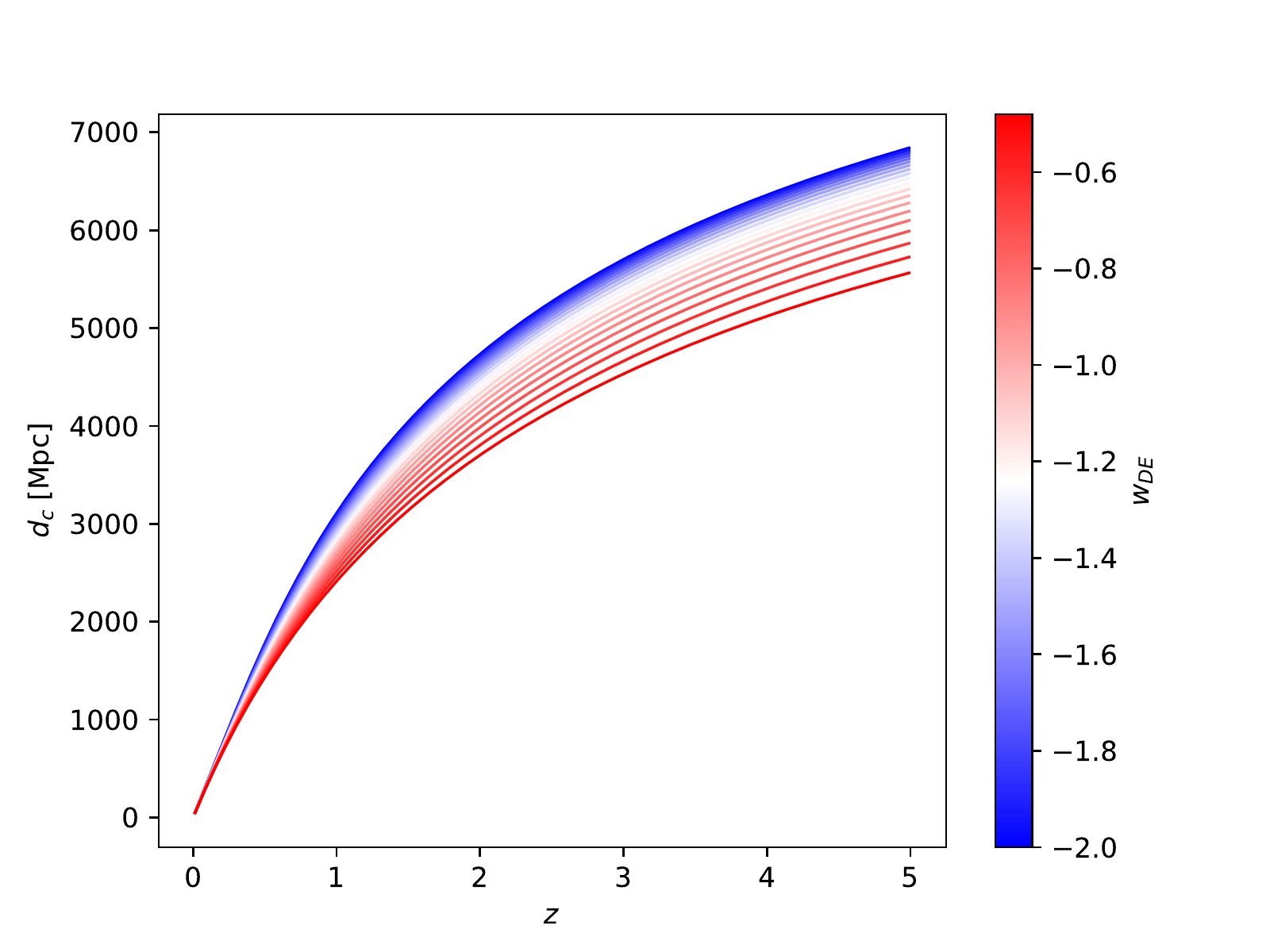}
\end{center}
	\caption{\label{fig:distance}Comoving distance for different values of the Hubble constant $H_0$ (upper panel), matter density parameter $\Omega_m$ (medium) and the DE equation of state $w_{DE}$ (lower). }
\end{figure}

Figure~\ref{fig:distance} shows the comoving distance for different values of the Hubble parameter $H_0$, the matter density parameter $\Omega_m$ and the dark energy (DE) equation of state $w_{DE} =$ constant. Note that these quantities are sensitive to the cosmological parameters and, therefore, observables related to the distance measurements are important to discriminate among the possible values of these parameters. Three of these observables are the distance modulus of type Ia supernovae (SNe Ia) \cite{Branch1993,Betoule2014}, the baryon acoustic oscillations (BAO) \cite{Bassett2010} and the so-called cosmic chronometers.

\subsection{Equations of motion}
\label{sec:eqm}

Up to this point we discussed the kinetic aspects of our choice of metric.
Nevertheless, we need a way to compute the time evolution of $a(t)$ in order to
calculate the quantities above. In this section we will discuss the equations of motion in the context of GR.

The standard cosmological model assumes GR and the FLRW on large-scales and
thus, the equations of motion are given by the so-called Friedmann equations,
\begin{align}
\label{eq:friedmann1} 
\left(\frac{\dot{a}}{a}\right)^2 &= \frac{8\pi G \rho}{3c^2} - \frac{Kc^2}{a^2}, \\
\label{eq:friedmann2}
\frac{\ddot{a}}{a} &= - \frac{4\pi G}{3c^2} \left( \rho + 3p\right),
\end{align}
where $G$ is the gravitational constant, $\rho$ and $p$ are the  energy density
and pressure of the matter-energy content of the universe, which is composed by
radiation (photons and relativistic neutrinos), matter (Dark Matter (DM),
baryons and neutrinos) and DE. Their equation of state, $p = w
\rho$,  are
\begin{align}\label{eq:w:r}
\text{Radiation}: w_r &= \frac{1}{3} \, , \\ \label{eq:w:m}
\text{Matter}: w_m &= 0 \, , \\  \label{eq:w:de}
\text{Dark Energy}: w_{DE} &\leqslant - \frac{1}{3} \, .
\end{align} 
The DE component of the standard model is, in general, described by the
cosmological constant $\Lambda$, in which $w_{DE} = -1$.

It is clear from the Friedmann equations that we need additional equations of
motion to close the system, i.e., we need equations of motion that determine
$\rho_i$ and $p_i$ (or $w_i$), where $i$ represent the possible components. If
the matter component is minimally coupled with the  theory of gravitation and
its energy-momentum tensor is ``conserved'' (satisfy $\nabla_\mu T_i^{\mu\nu} =
0$), the imposition that this tensor is compatible with a Friedmann metric
results in the following equation of motion for $\rho_i$:
\begin{equation}\label{eq:rhoi}
\dot{\rho}_i+3H(\rho_i+p_i) = 0.
\end{equation} 
There are two important lessons to get from the above equation. Firstly,
Eq.~\eqref{eq:rhoi} is not enough to close the system of equations and,
therefore, we still need a way to determine $p_i$. This occurs due to the fact
that Eq.~\eqref{eq:friedmann1} is a constraint equation (see
Sec.~\ref{subsec:int_back} for more details ) and, consequently,
Eq.~\eqref{eq:friedmann2} can be derived from Eqs.~\eqref{eq:friedmann1} and
\eqref{eq:rhoi}. One common example  to solve the system and obtain $\rho_i(t)$
is to consider simple components, as  discussed above, where $w_i$ is a
constant. We have different options for more complicated fluids.  For instance,
for a scalar field  $\rho_i$ and $p_i$ are functions of the field variable and
the equation of motion of the scalar field can be used to close the system (in
this case Eq.~\eqref{eq:rhoi} is redundant). For a statistical description one
can use the background compatible  Boltzmann equation to describe the
phase-space distribution $f_i(x^\mu,p^\mu)$ and obtain $\rho_i$ and $p_i$ from
it [which is the case of the equations of state
Eqs.~\eqref{eq:w:r}-\eqref{eq:w:de}]. Finally, one can model $p_i$ (or $w_i$)
directly using a phenomenological approach.

The second lesson refers to the choice of the time variable.  That is, if we rewrite Eq.~\eqref{eq:rhoi} using $a$ as a time variable,
we obtain
\begin{equation}\label{eq:dec}
\frac{\partial{\rho}_i}{\partial a}+\frac{3}{a}(\rho_i+p_i) = 0.
\end{equation}
This simple manipulation has an important consequence.  The factor $H$ that
couples the original equation with all other components was eliminated. This
means that, for the case where $p(\rho)$ is a known function, this equation can
be solved without knowing any of the other components present  in the system. This
is one of the reasons why this kind of modeling [$p_i(\rho_i)$, $w_i(a)$, \dots]
is so popular.  It allows to obtain a solution for $\rho_i$ independently of the
other matter components and/or the gravitational theory.

Note that the density is a function of the time, $\rho = \rho(t) = \rho_r(t) + \rho_m (t) + \rho_{DE}(t)$, and it is not feasible to use observational data in order to recover $\rho(t)$ as given in the present form. Therefore, similarly to the curvature parameter $K$ (see Sec.~\ref{sec:background}), we rewrite Eq.~\eqref{eq:friedmann1} in terms of the redshift $z$ and define the density parameters as
\begin{equation}
\Omega_i = \frac{\rho_i(z = 0)}{\rho_{\mathrm{crit},0}}, \quad \rho_{\mathrm{crit},0} = \frac{3c^2H_0^2}{8\pi G},
\end{equation}
where $i$ stands for the radiation (r), matter (m) and DE components.\footnote{Note that here we are defining $\Omega_i$ as a constant evaluated today, in the literature it is also frequently found the definition as a function of $z$, i.e., $\Omega_i(z) \equiv \rho_i(z)/\rho_{\mathrm{crit}}(z)$, where $\rho_{\mathrm{crit}}(z) = {3c^2H^2(z)}/{(8\pi G)}$.}
Thus, Eq.~\eqref{eq:friedmann1} now reads
\begin{align} 
\nonumber
E^2(z) &\equiv \left( \frac{H(z)}{H_0} \right)^2 \\ \label{eq:Ez}
&= \Omega_r (1+z)^4 + \Omega_m (1 + z)^3 + \Omega_K (1+z)^2 \\
&+ \Omega_{DE} \exp\left\{{3 \int_0^z \frac{[1 + w_{DE}\left(z^\prime\right)]}{1+z^\prime} \dd z^\prime}\right\}.  \nonumber
\end{align} 
Computing this equation at $z = 0$, we obtain the constraint
\begin{equation}
\label{eq:constrain_Ok}
\Omega_r + \Omega_m + \Omega_{DE} + \Omega_K = 1,
\end{equation}
which reduces by one the size of the parameter space to study the recent
expansion evolution of the universe. In short, in the context of GR and
homogeneous and isotropic metric, one needs to estimate the following set of
cosmological parameters: $\{H_0, \Omega_r, \Omega_m, \Omega_{DE}, w(z)\}$, where
$\Omega_K$ is then determined by Eq.~\eqref{eq:constrain_Ok}. Assuming a flat
universe, $\Omega_K = 0$, which is largely considered in the literature given
the CMB constraints~\cite{Hinshaw2013,PlanckCollaboration2015}, we end up with
the restriction $\Omega_{DE} = 1 - \Omega_r - \Omega_m$.

\subsection{Integrating and evaluating the background}
\label{subsec:int_back}

In the last sections we described the observables related to the background
model. In order to compute them we need to integrate this model and save the
results in a way that they can be used in statistical calculations.  Both
tasks are crucial in numerical cosmology since the evolution of the
perturbations relies on several evaluations of the background model at different
times and, in a statistical analysis, these quantities need to be recomputed at
different points of the parameter space. For these reasons the computation
and evaluation times are bottlenecks for this kind of analysis.

\subsubsection{Integration strategies}

There are different strategies to integrate the background equations of motion.
First of all we need to choose how to integrate the gravitational part. The
Friedmann equations comprehend  a dynamical equation \eqref{eq:friedmann2} and a
constraint equation \eqref{eq:friedmann1}. This means that, in principle, we
need to solve the dynamical equation and impose the constraint equation in our
initial conditions. In other words, to solve Eq.~\eqref{eq:friedmann2}, we need
the initial conditions for both $a$ and $\dot{a}$ but this choice must satisfy
Eq.~\eqref{eq:friedmann1}. It is worth noting that, since the choice of an initial
value for $a$ is arbitrary, Eq.~\eqref{eq:friedmann2} ends up determining the
initial conditions completely. Therefore, in this context, there are no
dynamical degrees of freedom left for the gravitational sector, i.e., the
imposition of homogeneous and isotropic hypersurfaces removes the dynamical
degrees of freedom. 

In practice we want to solve the Friedmann equations in the expansion branch ($H > 0$). For this reason we know \emph{a priori} the sign of $H$. Therefore, we can take the square root of Eq.~\eqref{eq:friedmann1},
\begin{equation}
\frac{\dot{a}}{a} = \sqrt{\frac{8\pi G \rho}{3c^2} - \frac{kc^2}{a^2}}, 
\end{equation}
and solve it instead of solving Eq.~\eqref{eq:friedmann2}. We can even go
further, if all matter components decouples from gravity, as we discussed in
Sec.~\ref{sec:eqm}, then we have an analytical solution for $H$ given by
Eq.~\eqref{eq:Ez} if we choose $z$ (or $a$) as our time variable.

It is useful to review the steps that allowed an analytical solution for $H$.
First,  all matter components decoupled from the system as we argued in
Eq.~\eqref{eq:dec}. For this reason we were able to find solutions for $\rho_i$
as functions of $a$ (or any other time algebraically related to $a$). Second, we
used the first order Friedmann equation~\eqref{eq:friedmann1} to determine
$H^2$. Finally, we assumed that the model will always be in the expansion branch
$H>0$ allowing then the determination of $H$ as a function of $a$. If any of
these conditions were missing, i.e., one component not decoupling or the
indetermination of $H$ sign, the solution would involve the whole system.

In the general purpose code, like CLASS or CAMB, one cannot assume that all
matter components decouples from $H$. This happens because they include the
possibility of modeling the dark energy as a scalar field and the scalar field
equation of motion does not decouples from $H$. For instance,  a scalar field
$\varphi$ with canonical kinetic term and arbitrary potential $V(\varphi)$
satisfies,
\begin{equation}\label{eq:scf}
\ddot{\varphi} + 3H\dot{\varphi} + \frac{\partial V}{\partial \varphi} = 0.
\end{equation}

In particular, CLASS chooses the conformal time in unit of mega-parsec
($\mathrm{Mpc}$) to integrate all quantities. We summarize the strategy
implemented in CLASS in the following. They first analytically integrate every
component  that decouples from the system, e.g., the cold dark matter and baryon
fluids, and implement the energy density and pressure related to these
quantities as functions of $a$. Moreover, they implement the energy density and
pressure of more complicate components in terms of their internal degrees of
freedom.  For instance, a scalar field from the above 
example has $$\rho_\varphi = \frac{\dot{\varphi}^2}{2} + V, \quad
p_\varphi = \frac{\dot{\varphi}^2}{2} - V,$$ where the internal degrees of
freedom are $\varphi$ and $\dot{\varphi}$. They use the label `A' to reefer to
the first set, the functions of $a$ and the internal degrees of freedom
[$\rho_\varphi(\varphi, \dot{\varphi})$ and $p_\varphi(\varphi, \dot{\varphi})$
in the last example], and the label `B' for all internal degrees of freedom
[$a(t)$, $\varphi(t)$, $\dot{\varphi}(t)$ in the last example].   All other
functionals of these quantities, such as the distances defined in
Sec.~\ref{sec:dist},  are  labeled as `C'.  Finally they implement the
equations of motion for the `B+C' set and integrate all variables with respect
to the chosen time variable $\eta$.

We have two options to perform the `B+C' integration: (i)  integrate all at
once or (ii)  to first integrate `B' and then use the respective results to integrate `C'. Some
possible advantages of integrating all variables `B+C' together are:
\begin{enumerate}
	\item simplicity, one needs to integrate a system of variables with respect to time just once;
	\item  less integration overhead, the integration software itself has a computational cost.  Each time it is used there are the costs of initialization, step computation and destruction.
\end{enumerate}
On the other hand, the possible shortcomings are:
\begin{enumerate}
	\item step sizes smaller than the  necessary.  When performing the integration, the
	Ordinary Differential Equations (ODE) solver evaluates the steps of all
	components being integrated and adapts the time step such that the error bounds
	are respected by all components. For this reason, including the `C' set in the
	integration can result in a larger set of steps for all quantities `B+C'; 
	
	\item
	lack of modular format, in some analysis just a few (or just one) observables
	from `C' are necessary. When performing all integration at once you have two
	options, integrate everything every time, even when they are not all necessary,
	or to create a set of flags that control which quantities should be integrated by
	branching (e.g., if-statements). Naturally, the if-clauses create an
	unnecessary overhead if they are placed inside the integration loop.
	Alternatively, if one decides to create a loop for every combination to avoid
	this overhead, then he/she will end-up with $2^n$ different loops (where $n$ is
	the number of variables in `C') which creates new problems such as code
	repetition and harder code maintenance.
\end{enumerate}

In CLASS they integrate first `B' and then use the results to compute `C'.
However, they do not integrate the variables in `C'.  They just get the resulting
knots from `B' integration and compute at the same knots the values of `C' , and
then add to a large background table enclosing every variable in `A+B+C'. This
table is then used to compute the background variables at any time using a
spline interpolation. This means that no error control were used to compute the
`A' and `C' variables, even though it was used to compute `B'. 

In NumCosmo, they also integrate `B' first, but the `C' set is handled
differently. First,  all variables that are algebraically related to each other are identified. For example, the distances discussed in
Sec.~\ref{sec:dist} can be computed from the comoving distance without any
additional integration. Then a minimal set of variables in `C' is identified and
for each one a different object is built. For instance, the ones related to
distances are included in the \cobj{NcDistance} object. This object, when used,
integrates the comoving distance using the results from `B' present in the
basic background model described by a \cobj{NcHICosmo} object. The integration
is done requiring the same error bounds as in `B' and a different spline is
created for the comoving distance, with different time intervals. 

At this point the main differences between CLASS and NumCosmo  are that the first
does not integrate `C', it simply interpolates them based on a fixed grid choice,
and does not have a modular structure for the computations of `C'. Nevertheless, the
non-modular design choice is understandable.  When CLASS was first conceptualized
it intended to be a Boltzmann solver, thus, it is natural to always integrate
all quantities in `C' that are needed to compute the time evolution of the
perturbations. But now, CLASS is slowly migrating to a general purpose code as the
cosmological basis for different numerical experiments usually performed by
MontePython~\cite{Audren2013}. At the same time, NumCosmo was designed from the
ground-up to be a modular general purpose library to handle different
cosmological computations.

\subsubsection{Evaluating the background}

As we previously described, the integration output is usually saved in memory such that it
can be used latter through interpolation. In principle it would be also possible
to just integrate everything necessary at once. This can work for a simple code,
e.g., if we just need the conformal distance at some predefined set of
redshifts. However, in many other cases this would lead to a very complicated
code. For example, when integrating perturbations, we need to integrate it for
different values of the mode $k$. This means that we would have to integrate the
background and all modes $k$ at the same time. Not only that would be complex (a
multi-purpose code written like this would require a huge number of branches to
accommodate the different code usage), but sometimes one does not know \emph{a priori} which modes $k$ need to be integrated.

For this reason a good interpolation method is a central part of any numerical
cosmology code. The most common approach is the use of splines, which avoids the
Runge phenomenon for interpolation with several knots. A spline is defined as a
piecewise polynomial interpolation where each interval is described by a
polynomial of order $k$ and the whole function is required to be $C^{k-1}$,
i.e., a function with $k-1$ continuous derivative, see~\cite{Boor2001} for a
detailed description of spline and its characteristics. It is clear that such
piecewise function has $(n - 1) (k + 1)$ degrees of freedom where $n$ is the
number of knots, imposing continuity up to the $k-1$ derivative gives us
$(n-2)k$ constrains (remember that the continuity is imposed only on the
internal knots), therefore, there are $(n - 1) (k + 1) - (n-2)k = n + k - 1$
degrees left to determine. In practice, we interpolate functions that we know
its values at the $n$ knots, still leaving us with $k-1$ degrees of freedom to
determine. 

The simplest choice is the linear spline ($k=1$), in this case there are no
extra degrees of freedom to determine, nonetheless, the resulting function is
not very smooth, it is actually only continuous ($C^0$), and the interpolation
error is proportional to $h^1\mathrm{max}\left\vert f^\prime(x)\right\vert$,
where $h$ is the largest distance between two adjacent knots and $f(x)$ is the
function being interpolated and ${}^\prime$ the derivative with respect to $x$.
Hence, a linear spline is appropriated only for really small $h$ (large number
of knots) or for functions with small first derivatives. When we move to larger
$k$ we end up with the problem of choosing $k$ and then determining the extra
$k-1$ degrees of freedom. The first problem is solved in a geometrical manner,
the cubic splines $k=3$ are the ones that minimize $\int_{a}^{b}\dd
x\left[f^{\prime\prime}(x)\right]^2$, where $a$ and $b$ are the endpoints of the
complete interval being interpolated. This means that the cubic interpolation
produces functions with small curvature that still matches $f(x)$ at the knots.
This is a reasonable requirement, we are usually interested in interpolating
functions that does knot wiggle strongly between knots.

Choosing the cubic spline we then need to fix the remaining $2$ degrees of
freedom. This is usually done by imposing boundary conditions on the
interpolation function $p(x)$. For example, one can impose a value for
$p^\prime(a)=f^\prime(a)$ and $p^\prime(b)=f^\prime(b)$, which requires the
knowledge of the derivative of $f(x)$ at the interval extremities. Since this
information is usually not available other approaches are necessary. The
so-called natural splines impose that the second derivatives of the
interpolating function to be zero at $a$ and $b$, i.e., $p^{\prime\prime}(a) =
p^{\prime\prime}(b) = 0$, this algorithm can be found at the GNU Scientific
Library (GSL)~\cite{GSLPC2010}. Nevertheless, the imposition of an arbitrary value for the second derivative results in a global interpolation error proportional to
$h^2$, instead of the original $h^4$. Another approach is to estimate the
derivative at the boundaries and use it to fix $p^\prime(a)$ and $p^\prime(b)$,
this is the approach followed by CLASS (at least up to the version 2.6),
however, here we need to find a procedure that will result in a
$\mathcal{O}(h^4)$ error bound. Currently, CLASS uses the forward/backward three
point difference method which as an error bound of only $\mathcal{O}(h^3)$ which
spoils the global $\mathcal{O}(h^4)$ error bound of the spline interpolation. To
keep this error bound it is necessary a higher order approximation for the
derivatives~\cite{Boor2001, Behforooz1979}. Finally, this can also be solved by
imposing an additional condition on the interpolating function, in the
not-a-knot procedure we impose that the third derivative is also continuous at
the second and one before last knots. This condition maintains the global
$\mathcal{O}(h^4)$ error bound and can be easily integrated in the tridiagonal
system used to determine the spline coefficients, see~\cite{Boor2001} and
\cite[ncm\_spline\_cubic\_notaknot.c]{Vitenti2012c} for an implementation.

It is also worth mentioning that there are other similar choices of
interpolation. For example, in the Akima interpolation \cite[pg. 42]{Boor2001}
one estimates the first derivative at each knot using a simple two-points
forward/backward finite difference method and then use them and the function
values to determine a cubic polynomial at each interval. The resulting
interpolation function is not a spline since it is only $C^1$ and the
interpolation error $\mathcal{O}(h^2)$ is worse than a not-a-knot spline,
nevertheless, it is a local algorithm since it depends only on the knots and
their nearest neighbors and also simpler and faster. On the other hand, the
difference in speed for a spline algorithm is usually irrelevant. Notably, the
LSST-DESC Core Cosmology Library
(CCL)
\cite{CCL2018} is using the Akima interpolation as its default method (up to version v1.0.0).

When using interpolation through piecewise functions we have an additional
computational cost when evaluating the function. Given an evaluation point $x$ we need to determine to which interval this point belongs to. This is usually
accomplished performing a binary search (see~\cite{Heineman2008}, for example),
which is in the worst case $\mathcal{O}(\log n)$, where $n$ is the number of
knots. Some libraries, GSL for example, also provide an accelerated spline, i.e., in a
nutshell it saves the interval of the last evaluation and tries it first in the
next one. The rational here is that one usually evaluates the spline in an
ascending/descending in small steps (for instance, when integrating). However,
this has some disadvantages. First, if the evaluation is not done in an
ascending/descending order, it becomes useless. Since it saves the last step
in memory and modifies it in every step, it cannot be used in a multi-threaded
environment in a simple way.\footnote{It would be necessary to save/read the
	last evaluation interval in a per-thread basis.} Finally, the determination of
the interval usually contributes very little in the computation time, thus, in
general, it is safer and simpler to not use this kind of optimization.

The last point about the evaluation of piecewise functions is the determination
of the knots. In most codes this is done through some heuristic algorithm, in most cases the programmer uses the fixed end-points $a$ and $b$ and simply chooses the knots with linear/logarithm spacing, namely, $$x_i = a + \frac{(b-a)}{n-1}i, \qquad x_i = \exp\left[\ln a + \frac{(\ln b-\ln a)}{n-1}i\right],$$ where $i$ ranges from $0$ to $n-1$ and $n$ is the number of knots (sometimes a combination of these two is used). This approach has several pitfalls, first it is not clear the relation between $n$ and the final interpolation error. Hence, the final user has to vary the value of $n$ until he/she gets the desired precision. In a complex code there are sometimes very large number of splines and, consequently, the user has to play with a large set of control variables (for example $n_j$ where $j$ labels the different splines in the code, the point where to change from linear to logarithm scale, \emph{etc}) to attain a certain precision. Actually, the problem is even worse, the user can choose all control parameters, but in practice one does that for a given model with one fixed set of parameters (in the best case scenario a large number of  parameter sets are used). In a statistical analysis the model is evaluated at different points of the parameter space and nothing guarantees that the precision at these points will remain the same as in the points where it was calibrated. This problem is magnified in large parameter spaces considering that, in this case, it is harder to check for precision in the whole parameter space. 

We will close this section discussing two different methods to determine the
knots in a way that is adaptive to the desired precision. However, this is
usually algorithm-specific and need to be dealt case by case. First, consider
the case where the function to be interpolated is the result of an integral or a
solution of an ODE system. When it comes from an integral, we have 
\begin{equation}\label{eq:exp1}
F(x) = \int_a^b\dd x f(x).
\end{equation}
It is easy to transform this problem in a one-dimensional ODE, i.e., 
\begin{equation}\label{exp:exp2}
F^\prime(x)=f(x), \qquad F(a) = 0.
\end{equation}
Now, the integral or the ODE system can be solved with the many available ODE
integrators (see for example the excellent library
SUNDIALS~\cite{Hindmarsh2005}). The ODE solvers adapt the step automatically in
order to obtain the desired precision, moreover the step procedure is based in a
polynomial interpolation. For these reasons we can use the same steps to define
the knots to be used in the interpolation. In other words, the ODE solver
computes the steps necessary to integrate a function such that inside these
steps the function is well approximated by a polynomial (given the required
tolerance), therefore, it is natural to use the same steps to interpolate the
function afterwards.

The second method is applicable when we have a function $f(x)$ with a high
computational cost. When this function needs to be evaluated several times, it
is more efficient to build a spline approximation first and then evaluate the
spline. As a rule of thumb, this is useful if the function $f(x)$ will be
evaluate more than $n$ times, where $n$ is the number of knots you need to create the
spline interpolation. The method consists in comparing the value of the function
$f(x)$ and its spline approximation $p(x)$ inside of each interval of $p(x)$,
i.e., if $p(x)$ is defined with $n$ knots we compute the difference $$e_i = \left\vert
p(x_{i+1/2})-f(x_{i+1/2})\right\vert,$$ for each $n-1$ intervals, where $x_{i+1/2} = (x_i+x_{i+1})/2$. For each interval where $e_i$ does not satisfy the tolerance requirements ($e_i < f(x_{i+1/2}) r + a$, where $r$ is the relative tolerance and $a$ the absolute tolerance), we update the spline adding this new point (creating two new intervals) and mark them as not OK (\textbf{NOK}). The intervals where the tolerance is satisfied are marked as \textbf{OK}. After the first iteration throughout all knots repeat the same procedure for all intervals marked \textbf{NOK}, then repeat until all intervals are marked \textbf{OK}. An implementation of this algorithm can be found at~\cite[ncm\_spline\_func.c]{Vitenti2012c}. Some variants of this algorithm are also useful, for instance, instead of using the midpoint $(x_i+x_{i+1})/2$ one can also use the log-midpoint $\exp\left[(\ln x_i+\ln{x_{i+1}})/2\right]$ for positive definite $x_i$.\footnote{When $x_i$ are not positive definite one can use a similar algorithm with $\ln x_i\to\text{arctanh}(x_i/s)$ and $\exp\to \tanh$, where the is a scale $s$ controls the transition between linear and log scale.}

We presented above two methods to control errors when computing interpolation
functions. We argue that this kind of error control based on a single tolerance
parameter is essential for the numerical tools aimed for cosmology and astrophysics.\footnote{Usually the relative tolerance controls the final error of the code while the absolute tolerance is application specific and, in the few cases where it is used, it serves to avoid excess in the tolerance. For example, if a given function $f$ is zero at a point $x$ and we want to compute its approximation $p$, the error control is $$\vert f(x)-p(x)\vert < r f(x) + a \to \vert p(x)\vert <  a.$$ So, without the absolute tolerance, the error control would never accept the approximation, unless it was a perfect approximation $p(x)=0$.} The sheer amount of different computations needed to evaluate the cosmological and astrophysical observables require complex codes which handles many different numerical techniques. For this reason well design local error controls are necessary to have a final error control based on the tolerance required by the observable. This is also crucial when evaluating non-standard models, in these cases the codes tend to be less checked and tested, and after all  they are usually only used by the group that developed it. This same problem is also present in standard models but in regions of the parametric space far from the current expectations. These regions tend to be much less tested and sometimes excluded from the allowed parameters range.  

\section{Linear Perturbations}
\label{sec:lin_pert}

On top of the background standard model discussed in Sec.~\ref{sec:background},
we have the perturbation theory with which we describe the evolution of Universe
from the initial primordial fluctuations to the structure formation and their
imprints on the observables we measure. In this section we will have a lightning
review of perturbation theory in cosmology (a full description can be found
in~\cite{Peter2009}) and then we explore a single component scalar perturbation
equation in order to exemplify the common numerical challenges involved in this
context.

First, we need to extend our assumptions about the space-time geometry. Lets
call the background metric components in Eq.~\eqref{eq:rw_metric} by
$g^{(0)}_{\mu\nu}$. Now we assume that the metric describing our space-time is
given by
\begin{equation}
\label{perturb}
g_{\mu\nu}=g^{(0)}_{\mu\nu}+h_{\mu\nu},
\end{equation}
where 
\begin{align}
\label{perturb-componentes}
h_{00} &= 2\phi \nonumber, \\
h_{0i} &= -aD_i B, \\
h_{ij} &= 2a^{2}(\psi\gamma_{ij}-D_i D_j E), \nonumber
\end{align}
where $D_i$ is the covariant derivative with respect to the spatial projection
of the background metric. Here we are assuming that the metric is described by a
Friedmann metric plus small deviations, all degrees of freedom but the scalars
are ignored, leaving us with the fields $(\phi,\, \psi,\, B,\, E)$. The physical
idea is that a Friedmann metric is a good approximation and all deviations from
it can be described by a small perturbation. There are several important
theoretical aspects of this description that we are not going to discuss here,
ranging from gauge dependency~\cite{Bardeen1980, Mukhanov1992, Vitenti2014},
size validity of the approximation~\cite{Vitenti2012, Pinto-Neto2014} and its
theory of initial conditions~\cite{Peter2009}. Instead, we focus only on the
numerical methods to solve the perturbation equations of motion and evaluating
the result.

In order to be compatible with our choice of metric we require that the energy
momentum tensor must also be split into a background plus perturbations, following
a similar decomposition that we made for the metric. This produces the following
scalar degrees of freedom $(\delta\rho,\, \delta p,\, V,\, \Pi)$, where
$\delta\rho$ and $\delta p$ are the perturbations of the total energy density
and pressure respectively, $V$ the velocity potential and $\Pi$ the anisotropic
stress.

Now we have a much more complicated problem than the background. There are eight
degrees of freedom to determine among metric and energy momentum tensor
perturbations. This problem can be simplified by writing the equations of motion
in terms of geometric quantities, i.e.,
\begin{align*}
D_ia^i &= -D^2\phi, & D_iD_j\sigma^{ij} &= \frac{2}{3}D^2D^2_K\sigma, & \sigma &\equiv B-\dot{E}+2HE, \\ 
\delta H &= \dot{\psi}+H\phi + \frac{D^2\sigma}{3}, & \delta\mathcal{R} &=-4D^2_K\psi.
\end{align*}
Here all variables are defined with respect to the background foliation, $D^2$
is the spatial Laplacian and $D^2_K \equiv D^2+{3K}/{a^2}$ [where $K$ is the
spatial curvature of the background metric~\eqref{eq:rw_metric}], $a^i$
represent the acceleration of the Hubble flow, $\sigma$ the shear potential for
the Hubble flow lines, $\delta H$ the perturbation on the Hubble function and
finally $\delta\mathcal{R}$ the perturbation on the spatial Ricci tensor. In terms of these variables the first order Einsteins equations are
\begin{align}\label{eq:def:friedpA}
3H\delta H +\frac{\delta \mathcal{R}}{4} &= \frac{\kappa}{2}\delta\rho, \\ \label{eq:def:friedpB}
3\delta H - D_K^2\sigma &= \frac{3\kappa(\rho+p)}{2}V, \\   \label{eq:def:friedpCST}
3\dot{\delta H} + 9H\delta H + 3\dot{H}\phi + D^2\phi + \frac{\delta\mathcal{R}}{4} &= -\frac{3\kappa}{2}\delta p, \\ \label{eq:def:friedpCSD}
\psi - \phi - \dot{\sigma} - H\sigma  &= \kappa\Pi,
\end{align}
where $\kappa \equiv 8\pi G$ and taking $c=1$. The energy-momentum conservation provides two additional equations
\begin{align}
\dot{\delta\rho} + \dot{\rho}\phi+3H(\delta\rho+\delta p)+(3\delta H+D^2V)(\rho+p) &= 0, \\
\dot{V} - \phi + \frac{ \dot{p}}{\rho+p}V+\frac{\delta p}{\rho+p} - \frac{2D^2_K\Pi}{3(\rho+p)}&= 0. 
\end{align}
However, it is easy to check that these two equations are not independent from
Einsteins equations. It is a straightforward exercise to show that they can be
obtained, respectively, by differentiating Eqs.~\eqref{eq:def:friedpA} and
\eqref{eq:def:friedpB} and combining with the
Eqs.~\eqref{eq:def:friedpA}--\eqref{eq:def:friedpCSD}. Thus, we have eight
variables and only four equations of motion. Note, however, that $E$ and $B$
do not appear explicitly in the equations of motion, actually they appear only
in the variable $\sigma$. This is an artifact from the gauge dependency of the
perturbations, but since these equations are invariant through spatial gauge
transformation, they are automatically written in terms of variables invariant
under this gauge transformation (for mode details see~\cite{Vitenti2014}, for
example). 

The gauge transformations with respect to the scalar degrees of freedom involve
two scalar variables, one representing the spatial transformations that we just
discussed and another generating time transformations. This means that, instead
of fixing the gauge choosing a gauge condition we wrote the equations of motion
in a gauge invariant way. Note that this is operationally equal to fix the
gauge, for example, had we fixed the spatial gauge freedom by choosing $B=0$,
instead of $\sigma$ appearing in
Eqs.~\eqref{eq:def:friedpA}--\eqref{eq:def:friedpCSD} we would have the $E$
variable. Following this approach, we fix the temporal gauge by rewriting the
system of equations using the gauge invariant variables
\begin{align}\label{eq:gi:f}
\overline{\delta\rho} &\equiv \delta\rho - \dot{\rho}\sigma,   &
\overline{\delta p} &\equiv \delta p - \dot{p}\sigma, & \overline{V} &\equiv V + \sigma, \\
\Psi &\equiv \psi - H\sigma, & \Phi &\equiv \phi + \dot{\sigma}, & \zeta &\equiv (1 - l)\Psi + H \overline{V}, 
\end{align}
where $l \equiv 2K/[a^2\kappa(\rho+p)]$ and we introduced an additional variable,
the Mukhanov-Sasaki variable ($\zeta$), that will be useful later. In terms of
these variables, we have the Einsteins equations recast to
\begin{align}
3H(\dot{\Psi}+H\Phi)-D^2_K\Psi &= \frac{\kappa}{2}\overline{\delta\rho}, \\ \label{eq:Psi2}
\dot{\Psi}+H\Phi &= \frac{\kappa(\rho+p)}{2}\overline{V}, \\ \label{eq:bla}
\left(\frac{\dd}{\dd t}+3H\right)\left(\dot{\Psi} + H\Phi\right) -\frac{\kappa(\rho+p)\Phi}{2} 
- D^2_K \left(\frac{\Psi-\Phi}{3}\right) &= -\frac{\kappa}{2}\overline{\delta p},\\ \label{eq:diffPP}
\Psi-\Phi &= \kappa\Pi,
\end{align}
and the energy-momentum tensor conservation to
\begin{align}
\dot{\overline{\delta\rho}} + \dot{\rho}\Phi+3H\left(\overline{\delta\rho}+\overline{\delta p}\right)+\left(3\dot{\Psi}+3H\Phi+D^2\overline{V}\right)(\rho+p) &= 0, \\
\dot{\overline{V}} - \Phi + \frac{ \dot{p}}{\rho+p}\overline{V}+\frac{\overline{\delta p}}{\rho+p} - \frac{2D^2_K\Pi}{3(\rho+p)}&= 0. 
\end{align}
Hence, we reduced the problem from eight to six variables  by using the gauge
dependency. Nonetheless, we still have only four independent equations of
motion. 

Similarly to what happens to the background, Einstein's equations and energy-momentum conservation do not provide all the dynamics necessary to describe the
perturbations. Naturally, the matter components have their particular degrees of
freedom which in turn determine $\delta\rho$, $V$, $\delta p$ and $\Pi$. There
are some codes specialized in solving these set of equations when the matter
content is described by a distribution and its evolution by Boltzmann equations,
including CAMB, CLASS. Here we are going to focus in a much simpler problem,
which nevertheless, displays the numerical difficulties found when solving the
equations of state for the perturbations.

We can simplify the problem by including two assumptions. First, the matter
content does not produce anisotropic stress, i.e., $\Pi = 0$. For the second assumption we first define the entropy perturbation as
\begin{equation}\label{eq:S}
S \equiv \delta p - c_s^2\delta\rho, \qquad c_s^2 \equiv \frac{\dot{p}}{\dot{\rho}}.
\end{equation}
This definition is convenient since it defines a gauge invariant variable. For instance, using Eqs.~\eqref{eq:gi:f} we have $S =
\overline{\delta p} - c_s^2\overline{\delta\rho}$. Thus, our second assumption is
simply that there is no entropy perturbation, i.e., $S = 0$. Note that this will
be true for any barotropic fluid [a fluid satisfying $p(\rho)$]. We can better understand why these assumptions simplify the system by computing the equations of motion for $\zeta$, differentiating it with respect to time and combining with the other equations. Doing so, we get
\begin{equation}\label{eq:zeta}
\dot{\zeta} = H\left(\frac{2c_s^2D^2\Psi}{\kappa(\rho+p)} + \frac{2D^2\Pi}{3(\rho+p)} - \frac{S}{\rho+p}\right),
\end{equation}  
and to close the system we combine Eq.~\eqref{eq:Psi2}  and Eq.~\eqref{eq:diffPP} to obtain
\begin{equation}\label{eqdotzeta}
\frac{\dd}{\dd t}\left(\frac{a^3D^2\Psi}{H}\right) = \frac{\kappa(\rho+p)a^3D^2\zeta}{2H^2}+\kappa a^3D^2\Pi.
\end{equation}
We now recast into a more familiar form. Defining 
\begin{equation}\label{eq:Pzz}
P_\zeta \equiv \frac{a^3D^2\Psi}{H}, \qquad z^2 \equiv \frac{\kappa(\rho+p)a^3}{2H^2c_s^2},
\end{equation}
the equations are
\begin{align}\label{eq:z1}
\dot{\zeta} &= \frac{P_\zeta}{z^2} + \frac{2HD^2\Pi}{3(\rho+p)} - \frac{HS}{\rho+p}, \\ \label{eq:z2}
\dot{P}_\zeta &= z^2c_s^2D^2\zeta + \kappa a^2D^2\Pi.
\end{align}
The pair of equations above reduce to a simple second order differential equation when $\Pi = 0 = S$, this is the simplification that we are looking for.

In retrospect, we started with eight variables and six equations of motion, of
the six only four equations of motion were independent. Then we used the gauge
dependency to reduce the system to six variables. Furthermore, we combined the
system of equations of motion to arrive at two equations
\eqref{eq:z1}--\eqref{eq:z2} and four variables $(\zeta,\, P_\zeta,\, S,\,
\Pi)$. It is also worth noting that this combination of variables $\zeta$ is not
arbitrary, it comes naturally from the Hamiltonian of the system when the
constraints are reduced~\cite{Vitenti2013, Falciano2013, Peter2016}. Finally, when we apply the assumptions of $S = \Pi = 0$ we get 
\begin{equation}\label{eq:MS}
\dot{\zeta}_k = \frac{P_{\zeta k}}{z^2}, \qquad \dot{P}_{\zeta k} = -z^2w^2\zeta_k,\quad \text{or} \quad \ddot{\zeta}_k + 2\frac{\dot{z}}{z}\dot{\zeta}_k+w^2\zeta_k = 0,
\end{equation}
an harmonic oscillator with time dependent mass $z^2$ and frequency $w \equiv
c_s k/a$, where $k/a$ is the square root of the Laplacian eigenvalue used in the
mode of the harmonic decomposition.\footnote{We can write $\zeta = \zeta_k
	\mathcal{Y}_k$ and $P_\zeta = P_{\zeta k} \mathcal{Y}_k$ , for an eigenfunction
	$\mathcal{Y}_k$ of $-D^2$ with eigenvalue $k^2/a^2$, i.e., $D^2\mathcal{Y}_k =
	-k^2/a^2\mathcal{Y}_k$.}

\subsection{Numerical solution}

Equations~\eqref{eq:MS} are a reduced and simplified version of the cosmological
perturbation equations of motion. Nevertheless, they exhibit many of the
numerical challenges also present in a more complicated scenario. For this
reason in this section we discuss the numerical approaches used to solve them in
different contexts, then we make a short discussion about the difficulties that
arrive in a more complicated scenario.

Equation~\eqref{eq:MS} has three important regimes. In order to understand them, it is easier to first rewrite the equations using the variable $v\equiv z\zeta$, producing the equation of motion
\begin{equation}\label{eq:msv}
\ddot{v}_k+\left(w^2-\frac{\ddot{z}}{z}\right)v_k = 0.
\end{equation}
A single fluid with constant equation of state (constant $p/\rho$) in flat
hypersurfaces $K=0$, has $z \propto a^{3/2}$. In this case, the potential appearing in the equation above is [see Eq.~\eqref{eq:Pzz}],
\begin{equation}
V_z \equiv \frac{\ddot{z}}{z} = \frac{3\ddot{a}}{2a} + \frac{3\dot{a}^2}{4a^2}.
\end{equation}
In other words, the potential is proportional to a combination of $\dot{H}$ and
$H^2$ and consequently to the Ricci scalar. This amounts to show that the
potential in this case produces a distance scale close to the Hubble radius
squared at time $t$, i.e., $[c / H(t)]^2$.\footnote{It is for this reason that
	some authors use the expressions super-/sub-Hubble scales, scales much
	larger/smaller than $c/H(t)$. We can also found the expressions
	super-/sub-horizon to refer to the same scales. Nevertheless, this nomenclature
	is based on the fact that for some simple models the horizon is also
	proportional to $c/H(t)$, which can be wrong and counter-intuitive in many cases
	and as such should be avoided.} Inspired by this we define the potential scale
$d_{V} \equiv 1/\sqrt{\vert V_Z\vert}$.  The regime I takes place when
$$\mathrm{I:}\qquad w=\frac{a\lambda}{c_s} \ll d_V, \qquad \lambda \equiv \frac{1}{k},$$ 
that is, when the mode physical wave-length $a\lambda$ (where $\lambda$ is the
conformal wave-length) over the sound speed  is much smaller than the potential
scale. The regime II happens when the physical wave-length is much larger than the potential scale $d_V$, i.e.,
$$\mathrm{II:}\qquad w=\frac{a\lambda}{c_s} \gg d_V, $$ 
and finally the regime III is characterized by 
$$\mathrm{III:}\qquad w=\frac{a\lambda}{c_s} \approx d_V. $$ 

Each one of these different regimes have a particular numerical approach to
handle them. Moreover, there are situations where it is necessary to put initial
conditions on I and evolve up to III and vice-versa. For example, in
inflationary or bouncing primordial cosmological models one begins with quantum
fields in vacuum state. In these cases the modes of interest are in regime I and
must be evolved into regime III. Now, Boltzmann codes start the evolution in the
radiation dominated expansion past, in this regime all modes of interest are in
regime III, some components evolve to regime I and some stop at regime II.  For
these reasons it is necessary to have tools to handle both cases.

For instance, regimes II and III can be solved using common codes of numerical integration since they do not present an oscillatory behavior as regime I. To deal with this last regime, it is worth to use the Wentzel--Kramers--Brillouin (WKB)
approximation, as we describe in the following.\footnote{For a complete exposition about this subject
	see~\cite{Bender1978}.}
Equation~\eqref{eq:msv} has an approximate solution written in
terms of the time-dependent coefficients and their derivatives. To understand the nature of this solution, lets first
assume that all coefficients are constants. In this case the solutions would be
\begin{equation}\label{eq:WKB1}
v^\mathrm{s}_k = A_\mathrm{s} \sin\left(\int\dd t\; w\right), \qquad v^\mathrm{c}_k = A_\mathrm{c} \cos\left(\int\dd t\;w\right),
\end{equation} 
and a general solution a linear combination of these two. Note that we wrote
$\int\dd t$ instead of $t-t_0$, such that we can use it as a tentative solution
for the time-dependent coefficients. Computing $\ddot{v}_k$ using these
tentative solutions results in
\begin{align}\label{eq:d2vs}
\ddot{v}_k^\mathrm{s} &= -\left(w^2 - \frac{\ddot{A}_\mathrm{s}}{A_\mathrm{s}}\right)v_k^\mathrm{s} + \left[\frac{\dd}{\dd t}\ln\left(A_\mathrm{s}\sqrt{w}\right)\right]2wA_\mathrm{s}\cos\left(\int\dd t\;w\right),	 \\	 \label{eq:d2vc}
\ddot{v}_k^\mathrm{c} &= - \left(w^2 - \frac{\ddot{A}_\mathrm{c}}{A_\mathrm{c}}\right)v_k^\mathrm{c} - \left[\frac{\dd}{\dd t}\ln\left(A_\mathrm{c}\sqrt{w}\right)\right]2wA_\mathrm{c}\sin\left(\int\dd t\;w\right).
\end{align}
Naturally, our solution is not exact when the coefficients are time dependent.
The expressions above match the leading term for $\ddot{v}_k$ in this phase
($w^2={c_s^2k^2}/{a^2}$), and the extra terms represent the error present in
this approximation. Moreover, if we consider the limit of our approximation
($k\to+\infty$), the error grows to infinity. For this reason we can use the
freedom in choosing the function $A_\mathrm{s,c}$ to remove the diverging term,
i.e., $A_\mathrm{s,c} = A_w \equiv 1/\sqrt{w}$. This choice has other advantage,
it removed the term that mixed solutions of different phases (it mixed the cosine 
and sine solutions). Using this choice for $A_\mathrm{s,c} = A_w$ the second
time derivatives are expressed by
\begin{align*}
\ddot{v}_k^\mathrm{s,c} &= -\left[\left(w^2 - \frac{\ddot{z}}{z}\right)- \frac{\ddot{A}_w}{A_w}+ \frac{\ddot{z}}{z}\right]v_k^\mathrm{s,c},
\end{align*}
where we wrote explicitly the error $-{\ddot{A}_w}/{A_w}+ {\ddot{z}}/{z}$, which is
an improvement, since now it does not diverge in the limit $k\to+\infty$, but it stays constant in this limit. 

We can further improve our approximation. Starting with the same functional forms in Eqs.~\eqref{eq:WKB1} but with an arbitrary time-dependent frequency $\nu$, we arrive at the same second derivatives in Eqs.~\eqref{eq:d2vs} and \eqref{eq:d2vc} but with $w\to\nu$. Then making the equivalent choice for $A_\mathrm{s,c}$ ($A_\mathrm{s,c} = A_\nu=1/\sqrt{\nu}$) we finally get
\begin{align*}
\ddot{v}_k^\mathrm{s,c} &= -\left[\left(w^2 - \frac{\ddot{z}}{z}\right)+\nu^2-w^2- \frac{\ddot{A}_\nu}{A_\nu}+ \frac{\ddot{z}}{z}\right]v_k^\mathrm{s,c}.
\end{align*}
We already know that the choice $\nu = w$ results in a reasonable approximation with a error $\mathcal{O}(k^0)$. Now if we can try to correct $\nu$ to improve the approximation, for example using $\nu = w + f_1/(2w)$, we get the error
\begin{equation}
E_1 = f_1 + \frac{f_1^2}{4w^2}- \frac{\ddot{A}_\nu}{A_\nu}+ \frac{\ddot{z}}{z} = f_1 - \frac{\ddot{A}_w}{A_w} + \frac{\ddot{z}}{z} + \mathcal{O}\left(k^{-2}\right).
\end{equation}
Then, if we choose $f_1 = {\ddot{A}_w}/{A_w} - {\ddot{z}}/{z}$, our error improve to $\mathcal{O}\left(k^{-2}\right)$. 

In the approximations above, we note that, for an error
$\mathcal{O}\left(k^0\right)$ we need only the background variables and, for a
$\mathcal{O}\left(k^{-2}\right)$, second derivatives of the background variables.
It is easy to see that the same pattern continues, i.e., for a
$\mathcal{O}\left(k^{-2n}\right)$ error we need the $2n$ derivatives of the
background variables. This points out the first numerical problem with WKB
approximation, the need for higher derivatives for better approximations. In
practice, the background variables are most of the time determined numerically
as we discussed in Sec.~\ref{subsec:int_back}. This poses a natural problem for
the WKB approximation since we would need to compute the derivative numerically
or to obtain it during the background integration using analytic expressions. The equations of motion themselves  can be used to compute derivatives from the
variables states. Nevertheless, both approaches are limited, numerical
differentiation usually results in increasing errors for higher order
derivatives, which limits the usage of higher order WKB approximation. The
analytical approach through equations of motions provide an easy way to compute
low order derivatives but for high order derivatives it becomes more and more
complex. The complicated expressions resulting from this analytical approach
must be treated carefully, large and complicated analytical expressions frequently
produce large cancellation errors when computed naively, see Sec.~\ref{sec:can}
for a discussion on cancellation errors.




\section{Cancellation errors}
\label{sec:can}

In this section we discuss the most common cancellation errors that happens in
numerical cosmology, for a more in-depth discussion see, for
example~\cite{Kincaid2009}. One common pitfall that plagues numerical
computations is the cancellation error. The problem is a natural consequence of
the finite precision of the computer representation of real numbers, the
Floating-Point (FP) numbers. In this representation a real number is decomposed
in base and exponent. For example in a decimal base is $1.2345 \times 10^{5}$,
where the base is $1.2345$ and the exponent is $5$. In this example the base
occupies $5$ decimal places, which defines the precision of our
number.\footnote{The precision is usually defined in binary basis, since it is
	that which are used in the computer representation. This precision does not
	translate to a fixed number of decimal places. For instance, in the IEEE 754
	standard, the common used double precision FP number has a 53 bits base and 11
	bits exponent, this basis roughly translates to $15.95$ decimal places.} Now,
the common arithmetic operations, multiplication, sum and division can produce
round-off errors, i.e., operating on truncated and rounded numbers produce a
different result than operating on the actual numbers and then truncating. For
example, adding $n$ positive numbers leads to a $n\epsilon$ roundoff error in
the final result, where $\epsilon$ is the machine epsilon, the roundoff unit,
the smallest number such that $1 + \epsilon \neq 1$ in the FP representation,
see~\cite{Kincaid2009} for more details. In modern implementations of $64$-bits
double precision FP number $\epsilon \approx 2\times 10^{-16}$. This means that,
when adding positive numbers the roundoff contributes to $n \times 2\times
10^{-16}$ error, usually much smaller than other sources of
errors.\footnote{With the exception of computations that involves a very large
	number of operations, $n$ must be of the order of trillions ($\propto 10^{12}$)
	to produce a $0.01\%$ error. } For this reason, for these operations the
roundoff errors can be, in most cases, safely ignored. However, the cancellation
error happens when we subtract two close numbers. Namely, using the example
above with five decimal digits in the base, if we have two real numbers
represented exactly by $x = 1.23400000$ and $y = 1.23456789$, their subtraction
is given by $\delta_{xy} \equiv y - x = 0.00056789$. Now, the FP representation
of these same numbers are $\bar{x} = 1.2340$ and $\bar{y} = 1.2346$ and their
subtraction $\bar{\delta}_{xy} = \bar{y} - \bar{x} = 0.0006$. Note that
$\bar{\delta}_{xy} = 6\times 10^{-4}$ has now only one significant digit, while
the correct result is $\bar{\delta}_{xy} = 5.6789\times 10^{-4}$. This is a very
serious problem, any operation done using $\bar{\delta}_{xy}$ will provide a
result with at most one significant digit. Going further, in many cases the two numbers should be equal in the FP representation such that the result is either zero or $\propto \bar{y}\epsilon$ if they differ by a rounding operation. 

Fortunately, there are many cases where the problem can be avoided. For instance, given the function $$f(x) = \sqrt{1+x^2} - 1,$$ when the computer calculates its value for a given FP value of x, it divides the task in several steps:\footnote{In many cases the order of the operations follow a restrict precedence rule dependent on the compiler/specification.} $$\bar{x} := x; \quad \bar{y} := \bar{x} \times \bar{x}; \quad \bar{y} := \bar{y} + 1.0; \quad \bar{y} := \sqrt{\bar{y}}; \quad \bar{w} = 1.0; \quad \bar{f} = \bar{y} - \bar{w};$$
where $:=$ represents the attribution of a variable, the bar variables the FP
representations and $\bar{f}$ the final result. If $x < \sqrt{\epsilon}$ then we
have that the final value of $\bar{y}$ is exactly $1.0$, consequently $\bar{f}$
will be either zero or $\propto\epsilon$ depending on rounding errors. This is a
catastrophic result, for $x < \sqrt{\epsilon}$ this function produces no
significant digits! To make matters worse, such expressions frequently appear
within more complicated calculations, e.g., integrating $\int_0^1\dd{}xf(x)g(x)$
for a numerically well behaved function $g(x)$ produces a result with all or
even no significant digits depending on where $g(x)$ peaks (to the left or to
the right of $x \approx \sqrt{\epsilon}$), the integration routines estimate the
error assuming that all digits are significant, so the final error estimate
produced by the integrator can be much lower than the actual error.
Nevertheless, there
is a simple solution for this problem. Multiplying the numerator and the denominator of $f(x)$ by $\sqrt{1+x^2} + 1$ gives $$f(x) = \frac{x^2}{\sqrt{1+x^2} + 1},$$ which is numerically well behaved for any $x$ inside the FP representation. This shows that in some cases a simple manipulation of the expressions is enough to cure the cancellation error.

A more subtle example is the spherical Bessel function of order one $$j_1(x) = \frac{\sin x-x\cos x}{x^2}.$$ In this representation in terms of trigonometric functions, the function is not well behaved for $x \ll 1$. For $x\lesssim\sqrt{\epsilon}$, $\overline{\sin}(\bar{x}) = x$ and $\overline{\cos}(\bar{x}) = 1.0$, again the result is either zero or $\propto \epsilon/x^2$, for really small $x$ the $x^2$ can underflow to zero and produce $0/0$ or $\epsilon/0$, represented as the FP \textsf{nan} (not a number) or \textsf{inf} (infinity). In this case, there is no simple trick to make this function well behaved. The solution is to split the computation in two cases $x < c$ and $x \geq c$ for a large enough cut value of $c$. For $x \geq c$ we can use the expression above to compute $j_1(x)$, for the other branch we use the Maclaurin Series, i.e.,
$$j_1(x) = \frac{x}{3}-\frac{x^3}{30} + \frac{x^5}{840} + \dots,$$
where the cut $c$ and the amount of terms in the series used depend on the FP
precision.

This last example is actually a real world case, in order to compute the top-hat filtering (see for example~\cite{Penna-Lima2014}) of the matter power spectrum, one needs to integrate the power spectrum in $k$ times $\left[j_1(kR)\right]^2$.


\funding{`This research received no external funding.}

\acknowledgments{S.  D.  P.  V.  would  like  to  thank  financial  support  from  the  PNPD/CAPES  (Programa  Nacional  de  P\'os-Doutorado/Capes, reference 88887.311171/2018-00).}

\conflictsofinterest{`The authors declare no conflict of interest.} 

\abbreviations{The following abbreviations are used in this manuscript:\\

\noindent 
\begin{tabular}{@{}ll}
GR & General Relativity\\
SNeIa & Type Ia Supernovae\\
ODE & Ordinary Differential Equation \\
CMB & Cosmic Microwave Background \\
LSS & Large Scale Structure \\
NumCosmo & Numerical Cosmology \\
CAMB & Code for Anisotropies in the Microwave Background \\
CLASS & The Cosmic Linear Anisotropy Solving System \\
MCMC & Markov Chain Monte Carlo \\
FLRW & Friedmann-Lema\^itre-Robertson-Walker \\
DM & Dark Matter \\
DE & Dark Energy \\
WKB & Wentzel–Kramers–Brillouin \\
FP & Floating-Point 
\end{tabular}}

\reftitle{References}


\externalbibliography{yes}
\bibliography{references8}



\end{document}